\documentclass[prl,twocolumn,showpacs,preprintnumbers,superscriptaddress,amsmath,amssymb]{revtex4-1}
\usepackage{amsmath}
\usepackage{natbib}
\usepackage{amssymb}
\usepackage{color}
\usepackage{graphicx}
\usepackage{dcolumn} 
\usepackage{bm} 
\usepackage{hyperref}
\usepackage{longtable}
\usepackage{gensymb}
\usepackage{siunitx}
\usepackage[version=4]{mhchem}
\usepackage{tabularx}
\usepackage[normalem]{ulem}

\begin{document}
\texttt{}

\title{Surface state at BaSnO$_3$ evidenced by angle-resolved photoemission spectroscopy and \textit{ab initio} calculations}

\author{Muntaser Naamneh} 
\affiliation{Department of Physics, Ben-Gurion University of the Negev, Beer-Sheva, 84105, Israel}
\affiliation{Photon Science Division, Paul Scherrer Institut, CH-5232 Villigen, Switzerland}
\author{Abhinav Prakash}
\affiliation{Department of Chemical Engineering and Materials Science, University of Minnesota, 421 Washington Ave., Minneapolis, MN 55455}
\author{Eduardo B. Guedes}
\affiliation{Photon Science Division, Paul Scherrer Institut, CH-5232 Villigen, Switzerland}
\author{W. H. Brito}
\affiliation{Departamento de Física, Universidade Federal de Minas Gerais, C.P. 702, 30123-970 Belo Horizonte, Minas Gerais, Brazil}
\author{Ming Shi}
\affiliation{Photon Science Division, Paul Scherrer Institut, CH-5232 Villigen, Switzerland}
\author{Nicholas C. Plumb}
\affiliation{Photon Science Division, Paul Scherrer Institut, CH-5232 Villigen, Switzerland}
\author{Bharat Jalan}
\affiliation{Department of Chemical Engineering and Materials Science, University of Minnesota, 421 Washington Ave., Minneapolis, MN 55455}
\author{Milan~Radovi\'{c}}
\affiliation{Photon Science Division, Paul Scherrer Institut, CH-5232 Villigen, Switzerland}


\date{\today}

\begin{abstract}
Perovskite alkaline earth stannates, such as BaSnO$_3$ and SrSnO$_3$, showing light transparency and high electrical conductivity (when doped), have become promising candidates for novel optoelectrical devices. Such devices are mostly based on hetero-structures and understanding of their electronic structure, which usually deviate from the bulk, is mandatory for exploring a full application potential. Employing angle resolved photoemission spectroscopy and \textit{ab initio} calculations we reveal the existence of a 2-dimensional metallic state at the  SnO$_2$-terminated surface of a 1\% La-doped BaSnO$_3$ thin film. The observed surface state is characterized by distinct carrier density and a smaller effective mass in comparison with the corresponding bulk values. The small surface effective mass of about 0.12$m_e$ can cause an improvement of the electrical conductivity of BSO based heterostructures.

\end{abstract}


\maketitle

Perovskite oxides, in general, have attracted tremendous interest over the last decade for their potential to harbor new phenomena. In addition, their heterostructures, surfaces, and interfaces show huge prospects for technological applications \cite{Hwang:2012,Wang:2016}. For instance, the two-dimensional electron gas (2DEG) in the SrTiO$_3$/LaAlO$_3$ heterostructure \cite{Ohtomo:2004} and also in bare SrTiO$_3$ surface after light illumination \cite{Meevasana:2011,Santander:2011,Plumb:2014} exhibit exotic properties including superconductivity \cite{Reyren:2007}, magnetism \cite{Bert:2011}, and high electron mobility \cite{Hu:2018}.  

Over the last few years, there has been a surge of interest in perovskite alkaline earth stannates such as BaSnO$_3$ (BSO) and SrSnO$_3$ (SSO). Due to light transparency properties combined with high electrical conductivity, they are promising candidates to realize novel optoelectrical devices. In particular, BSO crystals as wide band gap insulator ($\sim$3.0 -- 4.0 eV) has shown unusually high room temperature electron mobility of over 100 cm$^2$V$^{-1}$s$^{-1}$ \cite{Luo:2012} while doped bulk BSO  reaches the mobility of 320 cm$^2$V$^{-1}$s$^{-1}$ \cite{Kim:2012}.  
Furthermore, both BSO and SSO in thin-film form showed a further improvement in light transparency, electrical conductivity \cite{Prakash:2017,Kim:2018} and high electron mobility generally attributed to a small electron effective mass in the range of 0.2-0.4m$_e$ \cite{Seo:2014,Lebens:2016,Allen:2016,Niedermeier:2017,Krishnaswamy:2017}.  Significant theoretical and experimental efforts \cite{Lochocki:2018} have been made to understand these unique electronic properties. Band structure calculations, based on density functional theory (DFT), found small effective masses for conducting state of bulk BSO \cite{Krishnaswamy:2017}, caused by a substantial reduction in the electron-phonon scattering rate.


Although many of the stannates' bulk properties are now well understood, very little is known about their surface properties. Generally speaking, depicting their surface electronic structures is essential for understanding interfaces' properties in heterostructures wherever the stannates are a constituent. Moreover, it was suggested that stannates host an interfacial electron gas as they experience band bending near the surface \cite{Xiaofeng:2014}. Still, such a two dimensional (2D) state has not been directly observed yet. To address this question, we performed angle-resolved photoemission spectroscopy (ARPES) combined with \textit{ab initio} calculations on La-doped BaSnO$_3$ (LBSO). Our spectroscopy measurements reveal the existence of a 2-dimensional-like state at the surface of LBSO thin films with distinct properties from the bulk. Comparing the experimental data with the calculated band structure for bulk BSO and slabs, we find that the SnO$_2$-terminated surface hosts the surface state with a smaller effective mass than the bulk. 

\begin{figure}[htb]
	\includegraphics[width=1\columnwidth]{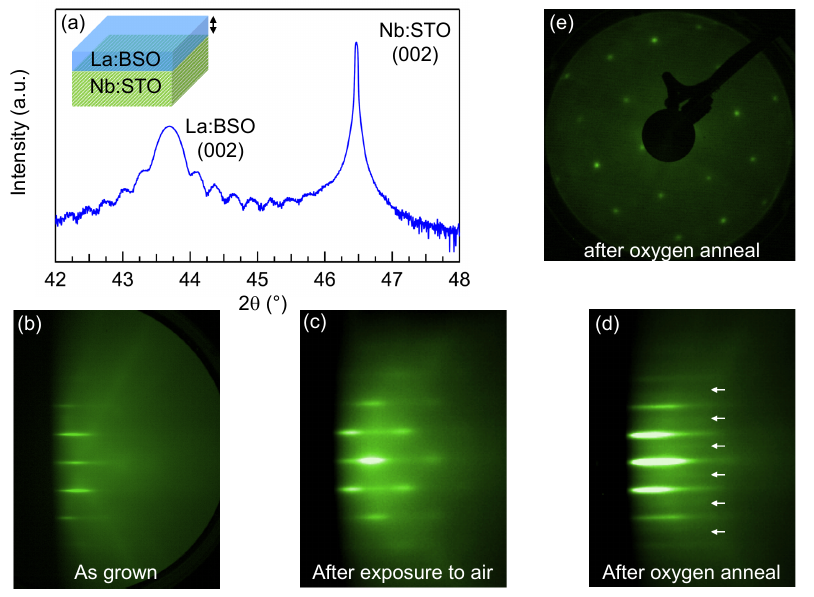} 
	\caption{(a) Wide-angle x-ray diffraction (WAXRD) of a 30 nm thick La-doped BaSnO$_3$ film grown on (001) oriented Nb-doped SrTiO$_3$ substrate. Reflection high energy electron diffraction pattern taken along [100] azimuth after (b) growth (in-situ at 14~keV), (c) ex-situ transfer and exposure to air and (d) high-temperature oxygen annealing to remove surface contamination. (e) Low-energy electron diffraction pattern from surface after oxygen annealing at 147 eV.}
	\label{fig1}
\end{figure}

In this Letter, we report the study of La-doped 30 nm thick BSO film grown by hybrid Molecular Beam Epitaxy (MBE)on Nb-doped STO (001) substrate. Here, we have chosen the conductive substrate to alleviate the charging effect during the subsequent photoemission experiments. Figure~\ref{fig1} (a) displays the wide-angle x-ray diffraction pattern of a LBSO film. The out-of-plane lattice parameter of 4.140 $\pm$ 0.002\AA~reveals that the studied LBSO film is mostly relaxed. A slightly higher out-of-plane lattice parameter is due to small residual biaxial strain in the film. The finite-size thickness fringes around the main x-ray diffraction (XRD) peak of the film (Fig.~\ref{fig1}~(a)) is indicative of a smooth film surface. 
The reflection high-energy electron diffraction (RHEED) patterns of the as-grown sample (measured at the University of Minnesota) and RHEED and low-energy electron diffraction (LEED) patterns after the ex-situ transfer (at the PLD-ARPES chamber at the SIS beamline in the Swiss Light Source) are shown in Figs.~\ref{fig1}~(b-e). The sample's RHEED pattern after transport was still well-defined, verifying the preservation of the surface's crystallinity. Though a relatively high background visible in the RHEED data indicates some surface contamination, and therefore we performed high-temperature annealing at 600\degree~C in an oxygen partial pressure of $1\times10^{-4}$~Torr. This procedure resulted in a significant improvement of the surface quality, as evidenced both by LEED and RHEED in Figs.~\ref{fig1}~(c)~and~(d). In addition, signatures of a (2$\times$1) reconstruction are visible (indicated by arrows in Fig.~\ref{fig1}~(f)), suggesting that the surface is well ordered.     


\begin{figure}[htb]
	\includegraphics[width=1\columnwidth]{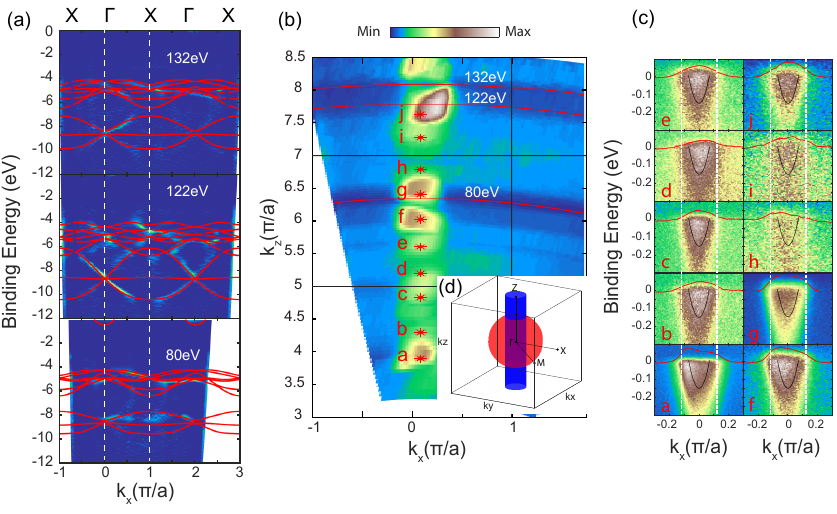} 
	\caption{(a) The valence bands dispersion along with the cuts parallel to $\Gamma$-X high symmetry direction overlaid with the linearized quasiparticle self-consistent GW band structure.  Cuts are acquired with incoming photon energies 80 eV, 122 eV, and 132 eV, as indicated by the panel's red lines (b). (b) The intensity map at Fermi energy in the k$_x$-k$_z$ plane was measured with incoming photon energy ranging from 20eV to 145eV. The black squares indicate the Brillouin zone boundaries informed from the periodicity of the valence bands. (c) The electron band near the Fermi energy obtained at different k$_z$ values indicated by red stars in panel (b). The sketch illustrates two possible Fermi surfaces in the momentum space. The 3-dimensional sphere-like FS for the bulk state is presented in red, and it is calculated using the nominal doping of the film. The 2-dimensional cylinder-like FS formed from the surface state is in blue, and its size corresponds to the measured one.}
	\label{fig2}
\end{figure}

ARPES is a powerful and unique tool to directly determine the electronic band structure of materials and its dimensionality. In Fig.~\ref{fig2}, we present the electronic band structure obtained in the k$_x$-k$_y$ plane. The k$_z$ momentum is perpendicular to the sample surface and thus orthogonal to the k$_x$ and k$_y$ momenta, lying on the surface plane along (100) and (010) cubic axis. Varying the incoming photon energy from 20~eV to 145~eV, we measured a band structure corresponding to different k$_z$ values. 
Fig.~\ref{fig2}(a) displays the deep valence band structure along with cuts parallel to high symmetry direction $\Gamma$-X passing through the different k$_z$ values. To better visualize the bands, the spectra have been processed using the 2D curvature method~\cite{Zhang:2011}.

In Figure~\ref{fig2}(a) we also compare the measured band structure to the linearized quasiparticle self-consistent GW (LQSGW) calculations for the bulk BSO using the experimental lattice constant of a = 4.14~\AA. For each spectrum image, we overlay the data with the calculated band structures at the corresponding k$_z$ value. The agreement between the experimental data and the calculations is overall very good, and a clear dispersion along with k$_z$, thus verifying their 3-dimensional nature. The calculated band gap of 3.4 eV also compares well to the experimental value of $\sim$3eV obtained from optical measurements \cite{Liu:2012,Seo:2014}. Further, we evaluate the conduction band effective masses in the $\Gamma$-X-direction, resulting in and 0.17m$_{e}$ . Our results are in good agreement with the values of 0.20 and 0.22m$_{e}$ reported in Refs.~\onlinecite{Krishnaswamy:2017,scanlon}, where the authors employed HSE06 and PBE0 hybrid functional, respectively, and with experimental values of 0.19m$_{e}$ reported in Ref.~\cite{Allen:2016}. In the Supplemental Material~\cite{suppl} we also present a comparison between LQSGW and DFT calculations, an expected smaller band gap of 1.4~eV was found, and an effective mass of 0.15m$_{e}$. The slightly smaller effective mass obtained within LQSGW indicate that electronic correlations have minor effects on the electron effective mass as expected for 5$s$ orbitals.






When the Ba atom is substituted with La, the conduction band, mainly composed of Sn(5s) orbitals, becomes occupied by the donated electrons turning LBSO into a metal. 
Previous studies showed that thin films of LBSO could exhibit high conductivity exceeding 10$^4$~Scm$^{-1}$, and it can be enhanced dramatically at the surface if it is interfaced with other insulators~\cite{Kim:2018}. Indeed, the properties of BSO based heterostructures strongly depend on the electronic structure of the interface region, which might bear similarities with the BSO film surface. Using low photon energies for ARPES ensures a short probing depth and thus high sensitivity to the surface electronic structure.

For the metallic bulk LBSO, the Fermi surface is expected to be 3-dimensional and isotropic in momentum space, with symmetry inherited from Sn(5s) orbital, as shown by the red sphere in the sketch of Fig.~\ref{fig2} (d). Nevertheless, the measured Fermi surface lying on the perpendicular k$_x$-k$_z$ plane at the Fermi energy (presented in Fig.~\ref{fig2} (b)) shows no clear dispersing bands along with the perpendicular momentum k$_z$. Occasionally such behavior can be attributed to the k$_z$ broadening effect, which occurs in ARPES measurements \cite{Lou:2018}, but this possibility is unlikely in this case. because the broadening effect would need to be enormous to stretch the expected small Fermi surface of LBSO to the Brillouin zone boundary. Besides, the variations in the intensity of the band at the Fermi energy do not follow the lattice periodicity, as indicated by the black squares in Fig.~\ref{fig2}~(b), and the states at the Fermi level reach the Brillouin zone boundary at Z point. Finally, Fermi momentum value and bandwidth, are similar at different k$_z$ values (see Fig.~\ref{fig2}~(c)). Therefore, one can see that the measured Fermi surface is rather cylindrical (sketched in blue in Fig.~\ref{fig2}~(d)) than spherical (shown by the red sphere), indicating the 2D nature of the measured band structure.

\begin{figure}[htb]
	\includegraphics[width=1\columnwidth]{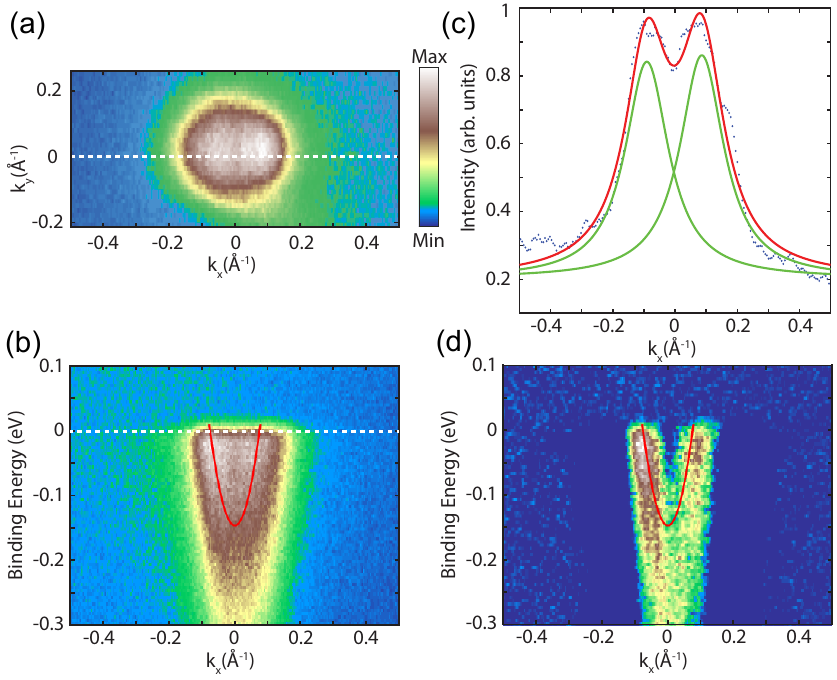} 
	\caption{The photoemission spectrum near the Fermi energy, which was acquired with an incoming photon energy of 28~eV. (a) The intensity map at Fermi energy was obtained from several cuts at different k$_y$ values. (b) Spectrum image obtained along cut parallel to the $\Gamma$-X direction as shown by the dashed white line in (a) showing the dispersion of the electron pocket along with k$_x$ with k$_y$=0. (c) Momentum distribution curve (MDC) extracted from panel (b) at E=E$_F$ and the fitted to two Lorentzian profiles to mark the position of k$_F$. (d) The second derivative along the energy axis for the spectrum image in (b) indicates the dispersing band.}
	\label{fig3}
\end{figure}

To obtain more quantitative information about this 2D-like band, we performed detailed ARPES measurement using a photon energy of 28~eV, which provides better energy and momentum resolutions among all used here. The constant energy intensity map at Fermi energy built from several cuts acquired along the direction parallel to $\Gamma$-X (see Fig.~\ref{fig3}~(a)), and confirms the circular shape formed from the electron-like band as shown both in Fig.~\ref{fig3}~(b), and in the second derivative analysis of Fig.~\ref{fig3} (d). Based on the Luttinger theorem \cite{Luttinger:1960}, we can estimate the carrier density, which is proportional to the enclosed volume by the Fermi surface. From the analysis of the momentum distribution curve (MDC) we extracted the value of k$_F$ by fitting the MDC at the Fermi energy to two Lorentzian curves (see Fig.~\ref{fig3} (c)).  
The measured k$_F$ of the 2D Fermi surface (k$_F\sim 0.067$~\AA$^{-1}$) amounts to a charge carrier density of n$_{2D}^{ARPES}$=1.2$\times$10$^{-2}$e$^-$/a$^2$. The carrier density measured by transport for this film \cite{Prakash:2017} is n$_{3D}$=1.7$\times$10$^{20}$1$^-$/cm$^3$ , which corresponds also to n$_{3D}$=1.2$\times$10$^{-2}$e$^-$/a$^3$. The similar electrons concentration per unit cell for the surface state and the bulk suggests homogeneous doping of La across whole film, and implies that the conduction states show an altered dimensionality at the surface. These surface states are reminiscent Shockley states since they have features of nearly free electron states and occur essentially due to SnO$_2$ termination \cite{Shockley:1939}.

Tracking the band dispersion in Fig.~\ref{fig3}~(b) and (d) reveals other unique properties of this 2D state. Keeping k$_F$ fixed, as determined by Fig.~\ref{fig3}~(c), the experimental spectrum can be well pictured by a parabola with the minimum around 0.15~eV and effective mass around 0.12~$m_e$. The observed value of $m^*$ is smaller than the previously reported values from bulk optical measurements \cite{Allen:2016} and predicted by bulk DFT calculations \cite{Lebens:2016}. Interestingly, the spectra presented in Fig.~\ref{fig3}, although with a moderate background, indicated that the band is very broad. This broadening can be a sign of many-body interactions, such as electron-phonon coupling. Indeed, it was reported that strong electron-phonon coupling of the conduction electrons with the LO$_{1,2,3}$ phonon modes \cite{NiedermeierPRB2017} is present in BSO. It is also important to note the strong incoherent spectral weight around the $\Gamma$-point is also found in low doped STO, which is related to the presence of strong electron-phonon coupling \cite{Wang:2016}.

The above results suggest the existence of a 2-dimensional state at the LBSO surface, which can be related to the observed 2DEG-like behavior of electrical properties in  LaInO$_3$/BaSnO$_3$ bilayer \cite{Kim:2018}. However, our ARPES experiment cannot solely reveal the origin of such state. In general, the breaking of translation symmetry on the surface may result in a new state whose wave function localized at the surface
~\cite{Shockley:1939}, and has different properties than the bulk band. Doping via oxygen vacancies, in cooperation with downward band bending, has been proposed to explain the formation of the 2-dimensional state in SrTiO$_3$ surfaces and interfaces \cite{Meevasana:2011,Santander:2011,Plumb:2014}, but this is not likely to be the case for BSO films based on the observation of an upward band bending by photoemission spectroscopy \cite{Lochocki:2018}.


\begin{figure}[htb]
	\includegraphics[width=1\columnwidth]{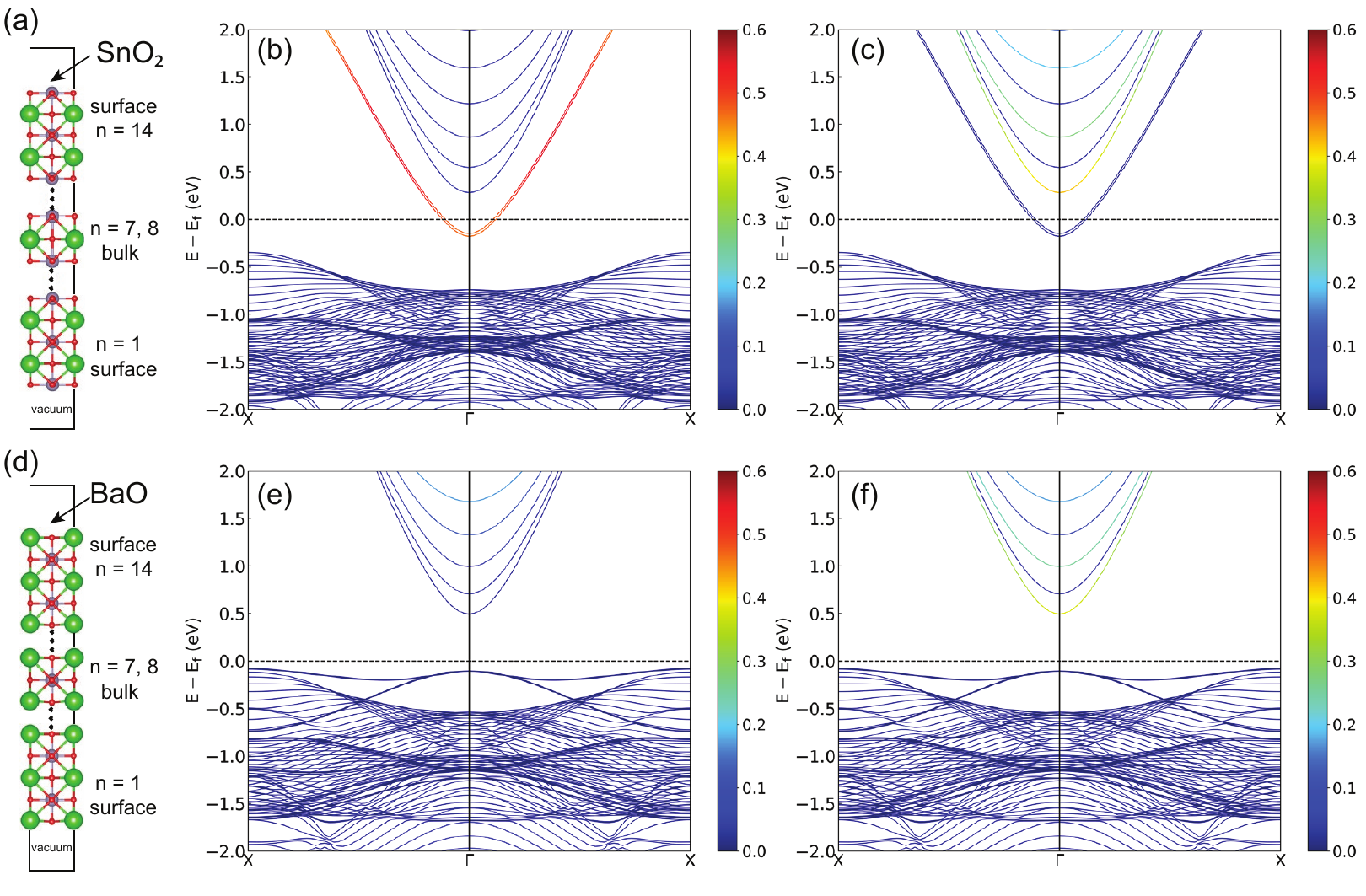} 
	\caption{(a) Optimized SnO$_2$-terminated slab structure with 14 SnO$_2$ layers. Sn, O, and Ba atoms are represented as grey, red and green spheres. (b) Orbital-resolved DFT slab band structures along X-$\Gamma$-X highlighting the contribution of Sn(5s) states from the surface and (c) from the bulk (central layer). (d), (e)  and (f) show the equivalent data for a BaO-terminated slab.}
	\label{fig4}
\end{figure}

To investigate the origin of the experimentally observed 2D states, we performed DFT(GGA-PBE) calculations for BSO slabs on (001) surface, with both SnO$_2$ and BaO terminations. Our optimized SnO$_2$-terminated BSO structure is shown in Fig.~\ref{fig4} (a). While the BSO bulk band structure (Fig.~\ref{fig2}) displays a clear bandgap, the resulting band structure for the SnO$_2$-terminated slab shows parabolic-like semi-occupied states. Figs.~\ref{fig4} (b) and (c) present the orbital resolved band structure highlighting the contributions of Sn-5s orbitals from both bulk and surface regions. Our calculations show that the states crossing the Fermi level are mainly derived from Sn(5s) orbitals from the surface atoms, while the Sn(5s) states from an atom in the bulk region of the slab appear around 0.48 eV above the surface states. In contrast, the band structure of the BaO-terminated slab (Fig.~\ref{fig4}~(d)) does not show any band crossing the Fermi level around the $\Gamma$ point, as seen in  Figs.~\ref{fig4} (e) and (f). For this termination, the Sn(5s) states of atoms just below the BaO surface present a small contribution at higher energies. In comparison, the bulk Sn(5s) states compose the majority of the bottom of the conduction band (Fig.~\ref{fig4} (f)). 

In addition, the calculated effective masses of both surface and bulk states of our BSO slab (Fig.~\ref{fig4} (b)) are close to 0.18 m$_{e}$, which is slightly smaller than the reported one \cite{Allen:2016}, but its is still larger than the observed value of 0.12$m_{e}$. The discrepancy between experimentally and theoretically obtained bulk values for the effective mass can be assigned to the existence of point defects in the sample or screening of scattering processes, causing enhancements of the electrical conductivity of BSO based heterostructures. However, the difference in the predicted and observed effective masses of the 2D state might have a more intricate origin, which can be related to a particular surface relaxation not captured by our calculations. The smaller surface effective mass can be attributed to a more considerable overlapping between $s$ orbitals, which can enhance electrical conductivity.





Combining angle-resolved photoemission spectroscopy (ARPES) with \textit{ab initio} calculations, we reveal the existence of a 2-dimensional metallic state at the surface of 1\% LBSO thin film. Comparing the experimental data with the calculated band structure for both bulk BSO and slabs, we found that only the SnO$_2$-terminated surface hosts the surface state. While our calculations describe the measured valence bands well, the surface state manifests a unique behavior, significantly different from the bulk. This surface state's particular property is its small effective mass of about 0.12~$m_e$, qualifying BSO as an excellent platform for optoelectronic devices. 


\section{Acknowledgments}
M.R. and E.B.G  were supported by SNSF Research Grant 200021\_182695. W.H.B. acknowledges the Pró-Reitoria de Pesquisa of Universidade Federal de Minas Gerais, and the National Laboratory for Scientific Computing (LNCC/MCTI, Brazil) for providing HPC resources of the SDumont supercomputer, which have contributed to the research results, URL: http://sdumont.lncc.br. The work at UMN was supported by the NSF DMR-1741801 and in part by the Air Force Office of Scientific Research (AFOSR) through Grant Nos. FA9550-19-1-0245 and FA9550-21-1-0025. Parts of this work were carried out at the Minnesota Nano Center and Characterization Facility, University of Minnesota, which receives partial support from NSF through the MRSEC program DMR-2011401

\pagenumbering{gobble}

\bibliography{refs.bib}
\end{document}


\texttt{}

\title{Supplementary information for ``Surface state at BaSnO$_3$ evidenced by angle-resolved photoemission spectroscopy and \textit{ab initio} calculations''}
\author{Muntaser Naamneh} 
\affiliation{Department of Physics, Ben-Gurion University of the Negev, Beer-Sheva, 84105, Israel}
\affiliation{Photon Science Division, Paul Scherrer Institut, CH-5232 Villigen, Switzerland}
\author{Abhinav Prakash}
\affiliation{Department of Chemical Engineering and Materials Science, University of Minnesota, 421 Washington Ave., Minneapolis, MN 55455}
\author{Eduardo B. Guedes}
\affiliation{Photon Science Division, Paul Scherrer Institut, CH-5232 Villigen, Switzerland}
\author{Walber H. de Brito}
\affiliation{Departamento de Física, Universidade Federal de Minas Gerais, C.P. 702, 30123-970 Belo Horizonte, Minas Gerais, Brazil}
\author{Ming Shi}
\affiliation{Photon Science Division, Paul Scherrer Institut, CH-5232 Villigen, Switzerland}
\author{Nicholas C. Plumb}
\affiliation{Photon Science Division, Paul Scherrer Institut, CH-5232 Villigen, Switzerland}
\author{Bharat Jalan}
\affiliation{Department of Chemical Engineering and Materials Science, University of Minnesota, 421 Washington Ave., Minneapolis, MN 55455}
\author{Milan~Radovi\'{c}}
\affiliation{Photon Science Division, Paul Scherrer Institut, CH-5232 Villigen, Switzerland}

\date{\today}


\maketitle

\section{Growth} Thin film of BSO doped with 1\% of lanthanum were grown using oxide radical-based hybrid molecular beam epitaxy (MBE) in the growth facility at the University of Minnesota. 

\section{ARPES}  
\begin{figure}[htb]
	\includegraphics[width=1\columnwidth]{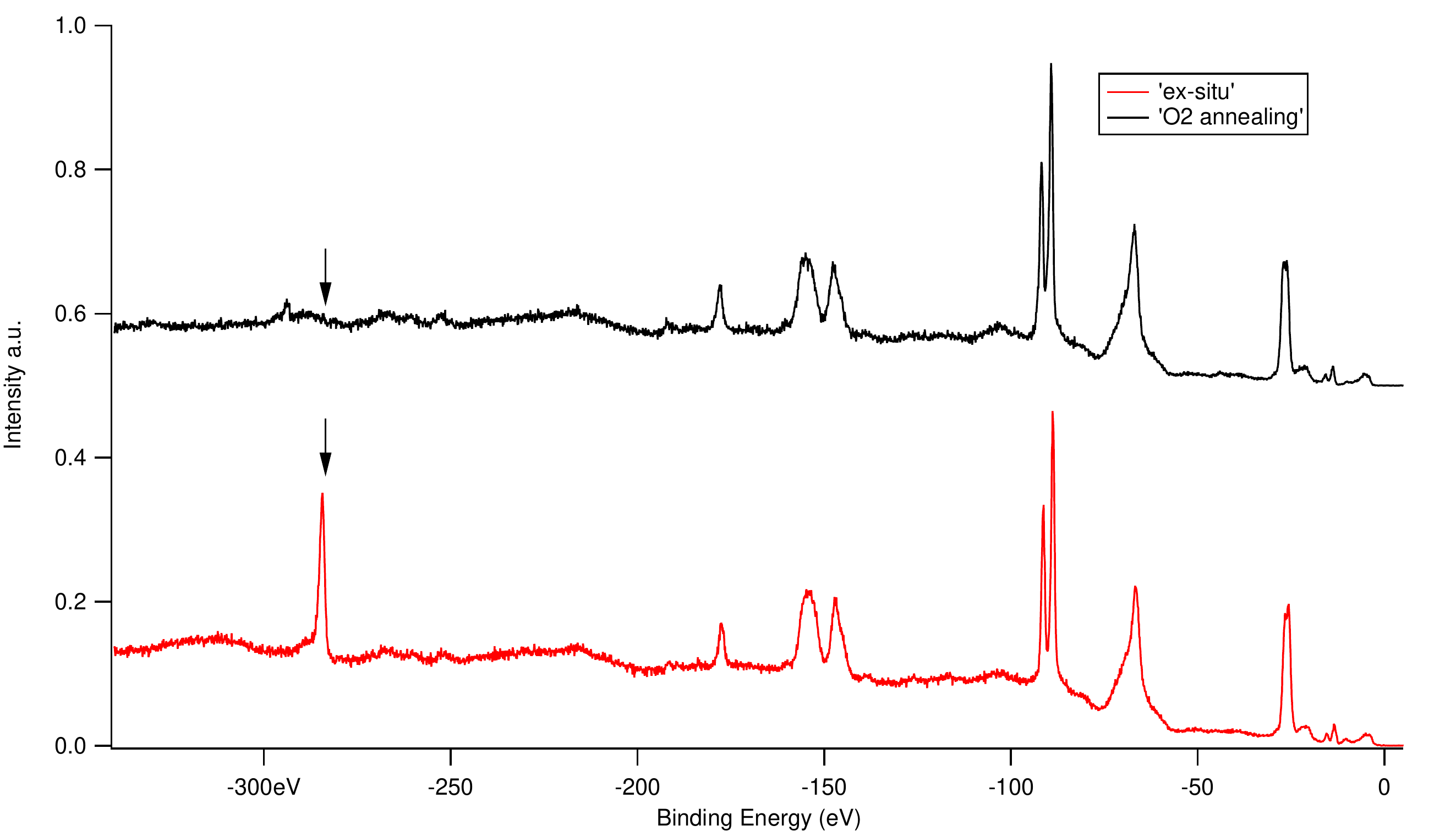} 
	\caption{XPS spectrum of the sample as introduced to the ARPES sample after ex-situ transfer, and XPS spectrum after annealing.}
	\label{figS1}
\end{figure}

Angle-resolved photoemission spectroscopy (ARPES) measurements were performed on the samples at the SIS beamline at Swiss Light Source after ex-situ transfer under ambient conditions. Reflection high-energy electron diffraction (RHEED) patterns were taken soon after the film growth (inside the MBE system at the Univ. of Minnesota) and after its reintroduction in the pulsed laser deposition (PLD)-ARPES system at the SIS beamline (Paul Scherrer Institute, Switzerland). The samples have also been characterized by x-ray photo-electron spectroscopy (XPS) before and after annealing as shown in Fig.~\ref{figS1}. The disappearance of the the peak indicated by the arrow indicate the removal of the contamination layer formed on the surface of the sample during the ex-situ transfer.

\begin{figure}[htb]
	\includegraphics[width=0.5\columnwidth]{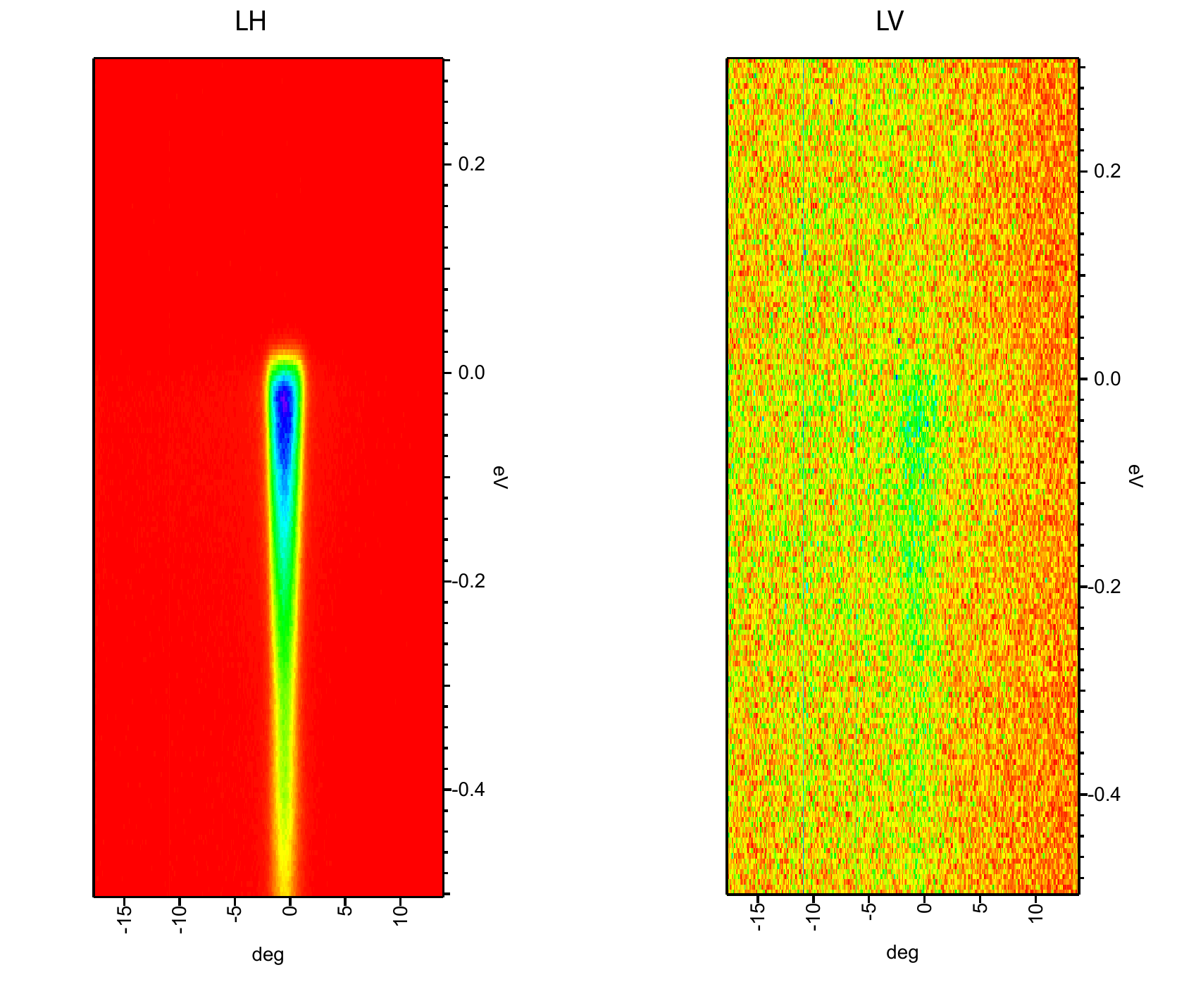} 
	\caption{The conduction band spectrum acquired with linear horizontal and linear vertical polarization.}
	\label{figS2}
\end{figure}

Polarization dependent ARPES measurements are powerful tool to identify the orbital symmetry in the electronic structure. The dipole selection rules can be exploited to determine the orbital symmetry with respect to a mirror plane of the crystal surface. In our experimental geometry, the dipole operator $A\cdot p$ has even parity with respect to the analyzer slit when the incident beam is horizontally polarized, and odd parity when the incident beam is vertically polarized. Regarding the final state, it is a plane wave with even parity with respect to the mirror plane. Therefore, the nonvanishing condition for the diploe transition $\braket{f|A\cdot p|i}$ is satisfied when the initial state has the same parity as the dipole operator. Therefore, the use of vertically polarized beam will cause the dipole transition to vanish for an s-orbital character and be visible for horizontally polarized beam as shown in Fig.~\ref{figS2}  

\section{Calculations} 
\subsection{Bulk}
Our DFT calculations for the bulk were performed within the Full Potential Linearized Augmented Plane Wave and localized orbitals (LAPW+lo) method, as implemented in the Wien2k package \cite{Madsen:2001,Blaha:2014}. In particular, we employed the PBEsol exchange-correlation potential \cite{Perdew:2008}. Our LQSGW calculations were done using the FLAPWMBPT code~\cite{kutepov1,kutepov2,kutepov3}  where the muffin-tin raddi in Bohr radius are 2.72, 2.43, and 1.46, for Ba, Sn, and O, respectively. 

\begin{figure}[htb]
	\includegraphics[width=0.7\columnwidth]{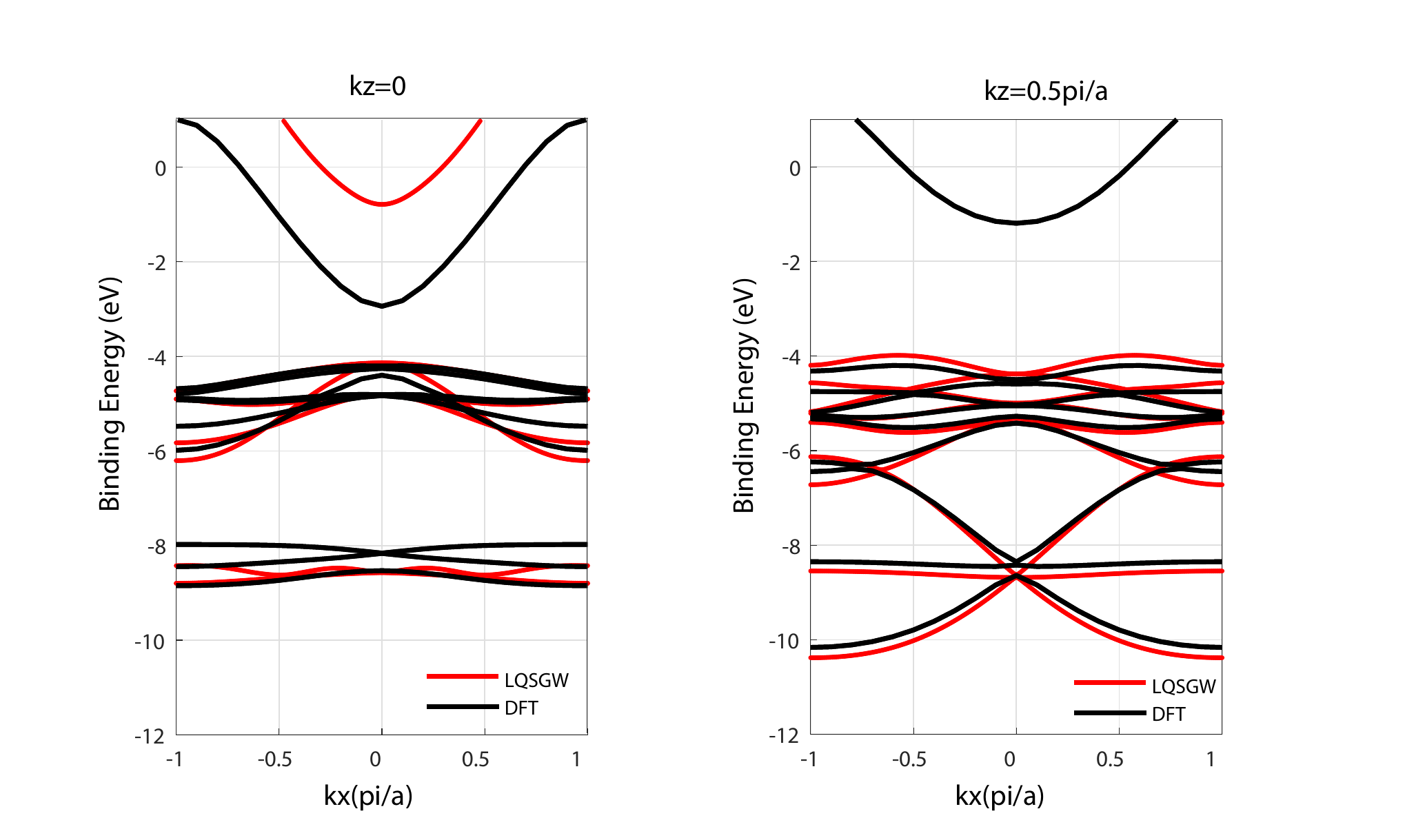} 
	\caption{DFT }
	\label{figS3}
\end{figure}

\subsection{Slab} The DFT calculations for the BSO slabs were carried out using the Vienna \textit{Ab initio} Simulation Package (VASP)~\cite{vasp1,vasp2}, where the wave functions were represented using the projector augmented wave (PAW) method~\cite{paw}.  We also used the Perdew-Burke-Ernzehof generalized gradient approximation (PBE-GGA)~\cite{pbe}, and a plane-wave basis with 520 eV cutoff. Atomic positions were relaxed until the forces on all atoms were reduced to 0.01 eV/\AA{}, using a 4 $\times$ 4 $\times$ 1 \textit{k}-mesh. Band structures were obtained using a 8 $\times$ 8 $\times$ 2 \textit{k}-points set.

\section{Process to find V$_0$} For each spectrum image, we overlay the data with the calculated band structures (both DFT and LQSGW) at different k$_z$ values until a good match was obtained. Following the agreement between the calculation and experimental band maps, and inferring from the periodicity of the valence bands measurements, we determined the inner potential to be 10eV.

\pagenumbering{gobble}

\bibliography{refs.bib}